\documentclass[showkeys,showpacs]{revtex4}
\usepackage{amsmath}
\usepackage{amssymb}
\usepackage{graphicx}

\begin{document}

\title{
Kaluza-Klein wormholes with the compactified fifth dimension
}

\author{
Vladimir Dzhunushaliev$^{1,2,3}$
\footnote{Email: v.dzhunushaliev@gmail.com}
and Vladimir Folomeev$^{2,3}$
\footnote{Email: vfolomeev@mail.ru}
}
\affiliation{$^1$Department of Theoretical and Nuclear Physics, Kazakh National
University, Almaty, 050040, Kazakhstan \\
$^2$Institut f\"ur Physik, Universit\"at Oldenburg, Postfach 2503,
D-26111 Oldenburg, Germany \\
$^3$Institute of Physicotechnical Problems and Material Science, National Academy of Sciences
of the Kyrgyz Republic, 265~a, Chui Street, Bishkek, 720071,  Kyrgyz Republic
}

\begin{abstract}
We consider wormhole solutions in five-dimensional Kaluza-Klein gravity
in the presence of a massless ghost four-dimensional scalar field.
The system possesses two types of topological nontriviality connected with the presence
of the scalar field and of a magnetic charge.
Mathematically, the presence of the charge appears in the fact that
the $S^3$ part of a spacetime metric is the
Hopf bundle $S^3 \rightarrow S^2$ with fibre $S^1$. We show that the
fifth dimension spanned on the sphere $S^1$ is compactified in the sense that
asymptotically, at large distances from the throat, the size of $S^1$ is equal to
some constant, the value of which can be chosen to lie, say, in the Planck region. Then, from the four-dimensional point of view,
such a wormhole contains a  radial magnetic (monopole) field, and an asymptotic four-dimensional
observer sees a wormhole with the compactified fifth dimension.
\end{abstract}

\maketitle

\section{Introduction}

Wormholes are hypothetical strongly gravitating
objects connecting two  (or more) distant regions of spacetime. At the present time, they have
several applications in theoretical physics, including:
(a) the attempt to describe physics as pure geometry (the model of a charged particle in the form of the Einstein-Rosen bridge~\cite{bridge});
(b) Wheeler's idea of ``charge without charge'' \cite{Misner:1957mt};
(c) the model of a traversable Lorentzian wormhole,
suggested by Morris and Thorne~\cite{morris} (for a review, see the book of Visser~\cite{visser});
(d) the development of the Euclidean approach to quantum gravity \cite{hawk}.

The principal feature of any wormhole is the presence of a nontrivial topology, and this appears in the fact that
such a configuration has a $n$-dimensional hypersurface of minimal  area. The possibility to construct such
solutions strongly depends on which particular model is under consideration.
For example, in four-dimensional Einstein's gravity (for which $n=2$)
static solutions describing the above-mentioned Lorentzian wormholes can be obtained only in the presence of the so-called exotic
matter violating the null energy condition (see, e.g., Ref.~\cite{visser}). In the case of modified gravity theories or
in considering wormhole models within the framework of multidimensional theories of gravity, the presence of exotic matter is not
already necessary (examples of such solutions can be found, e.g., in Refs.~\cite{wh_multidim}).

In the present paper we deal with  wormhole solutions within the framework of 5D Kaluza-Klein gravity.
The literature in the field offers simple analytical vacuum solutions, some of which may be
interpreted as describing Lorentzian wormholes \cite{Chodos:1980df,Angus:1985gb,Dzhunushaliev:1998ya,Chen:2000yi}.
These models contain three types of charges -- electric, magnetic, and scalar, the relation between which determines the
properties of resulting solutions.
Namely, one can obtain solutions describing spacetimes with trivial and nontrivial topologies, or they can contain
singularities, or, being regular, they can have a peculiar behavior in some regions of spacetime. The latter
is manifest as follows: In the case of wormhole-like solutions with only electric charge $q_e$,
the time coordinate changes its sign and becomes spacelike in inner regions of a configuration at  distances
$r^2<q_e^2$~\cite{Dzhunushaliev:1999rq}. Such a wormhole is nontraversable  in the sense that a test
particle cannot pass through it  from one asymptotically flat region to the other \cite{AzregAinou:1990zp}.
In the case where there is only a magnetic charge $q_m$, wormhole-like solutions have a true singularity at the points
 $r=q_m/2$ \cite{Dzhunushaliev:1998ya,Chen:2000yi}, and such solutions cannot in general be regarded as describing a wormhole
in the sense that they cannot be extended analytically to infinity. It is obvious that systems containing
both electric and magnetic charges will possess properties either of electric or of magnetic wormholes,
according as which of the fields dominates~\cite{Dzhunushaliev:1998ya,Chen:2000yi}.

The objective of the present paper is to construct static wormhole solutions with only magnetic charge which
(a)~are regular in the entire range $(-\infty <r <+\infty)$, (b) do not change the metric's signature,
(c) provide the size of an extra dimension to be of the order of the Planck length (since such configurations are of most interest
from the point of view of a four-dimensional observer).
To obtain solutions possessing such properties, here we will consider a five-dimensional Kaluza-Klein
wormhole in the presence of an additional ghost/phantom four-dimensional scalar field,
i.e., a scalar field with the opposite sign in front of the kinetic energy term. At the present time, such fields are widely used
in modeling the current accelerated expansion of the Universe (see, e.g., the reviews \cite{de_revs}).
 Their important property is that they may violate the weak/null energy conditions, thus providing the conditions  for
the creation of configurations with nontrivial topology.

The system will possess two types of topological nontriviality. The former is connected with the presence
of the ghost scalar field, which
determines a
wormhole topology of the system in question. The second topological nontriviality is due to the presence of a magnetic charge,
that results in the fact that the $S^3$ part of a spacetime metric is the Hopf bundle $S^3 \rightarrow S^2$~\cite{Steenrod}
with fibre $S^1$ where the fifth dimension is spanned on.
We will show below that in the model under consideration there is a possibility
of compactifying this fifth dimension to sizes, say, of the order of the Planck length. Then, from the point of view of a
four-dimensional observer, such a
wormhole may be regarded as an usual four-dimensional wormhole containing a radial magnetic (monopole) field.

\section{General 5D equations and their numerical solutions}
\label{5D_wh_sols}

As pointed out in the Introduction,
the objective of the present paper is to construct five-dimensional
wormhole solutions. To do this, we  consider a model of a gravitating massless
scalar field $\phi$ within the framework of  5D Kaluza-Klein gravity.  Our starting
point is the Lagrangian
\begin{equation}
	\mathcal L = - \frac{1}{2 \varkappa} R^{(5)} +
	\frac{\varepsilon}{2} \partial_\mu \phi \partial^\mu \phi,
\label{1-10}
\end{equation}
where $\varepsilon = \pm 1$ corresponds to usual and ghost scalar field, respectively, and
$\varkappa=8\pi \hat{G}$ with $\hat{G}$ to be  a  five-dimensional  gravitational  constant
(hereafter  we  work  in  units  such  that  $c=1$).
In considering wormhole-like configurations,  we take a static metric of the form
\begin{equation}
	ds^2 = dt^2 - dr^2 - a \left(
		d \theta ^2 + \sin^2 \theta d \varphi^2
	\right) - b^2 \left(
		d \psi + m \cos \theta d\varphi
	\right)^2,
\label{1-50}
\end{equation}
where the metric functions $a, b$ depend on the radial coordinate $r$ only,
and $\psi, \theta, \varphi$ are angular coordinates. The constant $m$ may be regarded as
a magnetic charge or, in mathematical language, as the Hopf invariant~\cite{Steenrod} (see below in Sec.~\ref{wh_prop}).

Because of the presence of nondiagonal components in the metric \eqref{1-50},
it is convenient to use the vielbein formalism, in which the Einstein equations are
\begin{equation}
	R^c_d - \frac{1}{2} \delta^c_d R = \varkappa T^c_d~,
\label{1-20}
\end{equation}
where $c,d=\bar{0}, \bar{1}, \bar{2}, \bar{3}, \bar{5}$ are the vielbein indexes. The corresponding energy-momentum tensor is
\begin{equation}
	T^\mu_\nu = \varepsilon\left(\partial_\nu \phi \partial^\mu \phi -
	\frac{1}{2}\, \delta^\mu_\nu \partial_\eta \phi \partial^\eta \phi\right),
\label{1-40}
\end{equation}
where
$\mu, \nu, \eta=0, 1, 2, 3, 5$ are the world indexes, and the conversion from the vielbein language to the language of world indexes is
performed in the usual manner. In turn,
a general equation for the scalar field is obtained by varying the Lagrangian~\eqref{1-10} with respect to $\phi$,
\begin{equation}
	\frac{1}{\sqrt{-g}} \frac{\partial}{\partial x^\mu} \left(
		\sqrt{-g} g^{\mu \nu} \frac{\partial \phi}{\partial x^\nu}
	\right)	= 0.
\label{1-30}
\end{equation}

Using the metric \eqref{1-50} and the energy-momentum tensor \eqref{1-40}, the
Einstein equations take the form
\begin{eqnarray}
	 \frac{a''}{a} - \frac{1}{4}\left(\frac{a'}{a}\right)^2 - \frac{1}{a} +
	 \frac{3}{4}\, m^2 \left(\frac{b}{a}\right)^2 &=&
	 - \frac{\varepsilon}{2} \frac{Q^2}{a^2 b^2}~,
\label{1-60}\\
	\frac{b''}{b} + \frac{a'}{a}\frac{b'}{b} -  \frac{1}{2}\, m^2 \left(\frac{b}{a}\right)^2 &=& 0~,
\label{1-70}\\
	- \frac{1}{4}\left(\frac{a'}{a}\right)^2  - \frac{a'}{a}\frac{b'}{b} + \frac{1}{a}
	-  \frac{1}{4}\, m^2 \left(\frac{b}{a}\right)^2 &=& - \frac{\varepsilon}{2} \frac{Q^2}{a^2 b^2}~,
\label{1-80}
\end{eqnarray}
where we have employed the solution of the scalar-field equation \eqref{1-30}
\begin{equation}
	\phi'(r) = \frac{Q_\phi}{a b}
\label{1-90}
\end{equation}
with the integration constant $Q_\phi $ (the scalar charge).
In the above equations, the prime denotes  differentiation with respect to the radial coordinate, and
$Q^2 = \varkappa\, Q_\phi^2$.
Equation \eqref{1-60} and \eqref{1-70} are the linear combinations
$2\left[ \binom{\bar{0}}{\bar{0}} - \binom{\bar{2}}{\bar{2}}\right]  - \binom{\bar{1}}{\bar{1}}$  and
$\left[ \binom{\bar{5}}{\bar{5}} - \binom{\bar{0}}{\bar{0}}\right]$ of the
components of the Einstein equations, and Eq.~\eqref{1-80} is the $\binom{\bar{1}}{\bar{1}}$ component.

Since here we seek wormhole solutions which are symmetric with respect to the center $r=0$,
the functions $a(r), b(r)$ must be even, and consequently
 $a'(0) = b'(0) = 0$. Then the constraint equation~\eqref{1-80} yields
\begin{equation}
	a(0) = \frac{m^2}{4}b^2(0) - \frac{\varepsilon}{2} \frac{Q^2}{b^2(0)}.
\label{1-100}
\end{equation}

Consideration of such symmetric solutions assumes that the parameter $\varepsilon$ must be taken as -1.
Indeed, by considering the $\left[ \binom{\bar{0}}{\bar{0}} - \binom{\bar{1}}{\bar{1}}\right] $ component of the Einstein equations
$$
\frac{a''}{a}+\frac{b''}{b}-\frac{1}{2}\left(\frac{a'}{a}\right)^2=-\varepsilon \frac{Q^2}{a^2 b^2},
$$
one can see that positive values of the second derivatives $a''(0), b''(0)$ corresponding
to a minimum of the functions $a, b$  at the throat can be realized only at $\varepsilon=-1$.


Let us now turn to a consideration of possible solutions to the system under consideration.
Consider first the case with $m=0$ that corresponds to the topologically trivial case with the Hopf invariant equal to zero.
(Notice that in this case the spacetime topology remains nontrivial.)
Then a solution to Eqs.~\eqref{1-60} and \eqref{1-70} is given by the expressions
\begin{eqnarray}
	a(r) &=& r^2 + \frac{Q^2}{2 b_c^2},
\label{1-104}\\
	b(r) &=& b_c = \mathrm{const}.
\label{1-108}
\end{eqnarray}

The expression \eqref{1-104} resembles the well-known 4D solution for a wormhole supported by a massless ghost
scalar field~\cite{wh_ghost}.
On the other hand, the 5D case being considered here contains the arbitrary
constant $b_c$ which can be chosen to have, for example,
the Planck values $b_c \sim {\cal O}(l_{\text{Pl}})$.
Thus the spacelike hypersurface  $t,r= \text{constant}$ has the $S^2 \times S^1$ topology,
and for a four-dimensional observer the solutions \eqref{1-104}, \eqref{1-108}
can be interpreted as describing a 4D wormhole embedded in a 5D spacetime with the compactified fifth dimension.

\begin{figure}[t]
\begin{minipage}[t]{.49\linewidth}
  \begin{center}
    \includegraphics[width=10.cm]{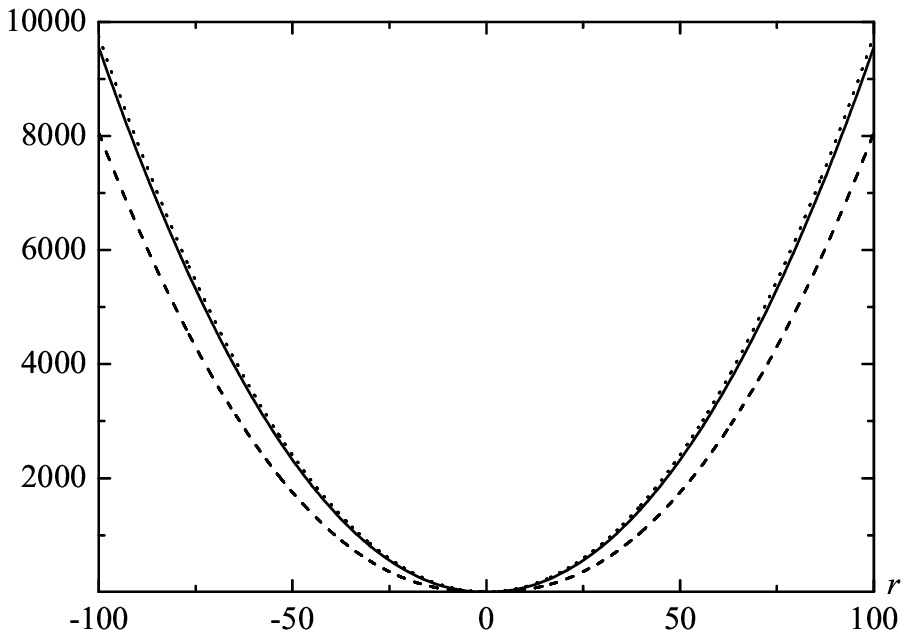}
  \end{center}
\end{minipage}\hfill
\begin{minipage}[t]{.49\linewidth}
  \begin{center}
  \includegraphics[width=9.4cm]{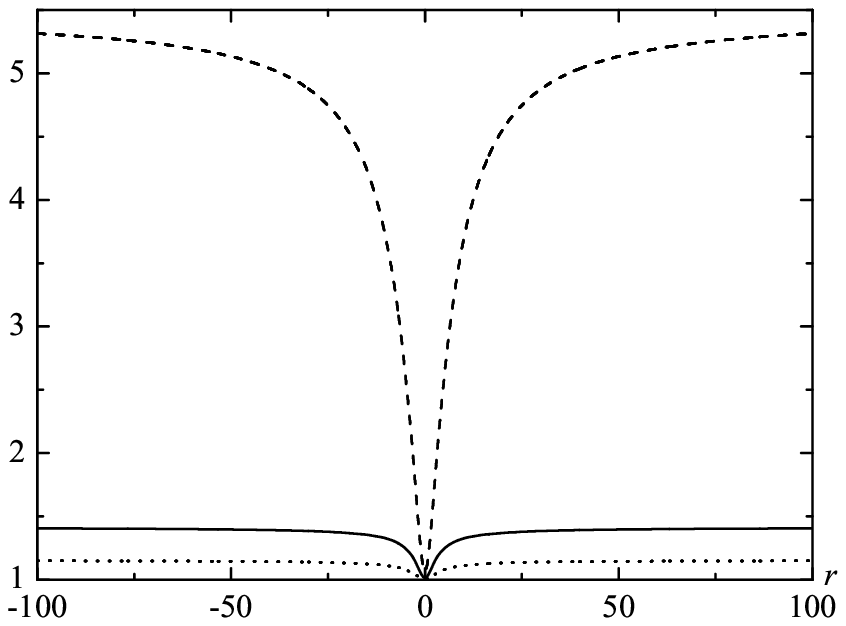}
  \end{center}
\end{minipage}\hfill
\vspace{-1.cm}
  \caption{The metric functions $a(r)$ (left panel) and $b(r)$ (right panel) are shown
  for different values of the scalar charge
  $Q_1 < Q_2  < Q_3$. The dashed lines correspond to $Q_1$, the solid lines - to $Q_2$, and the dotted lines - to $Q_3$. For both
  panels $m=1$.}
 \label{arbr}
\end{figure}

Consider now  topologically nontrivial solutions with a non-zero Hopf invariant $m$. In this case we will seek numerical solutions
to Eqs.~\eqref{1-60} and \eqref{1-70} using the following dimensionless variables:
\begin{equation}
\tilde{a} = \frac{a}{b^2(0)}, \quad
\tilde{b} = \frac{b}{b(0)}, \quad
\tilde{Q} = \frac{Q}{b^2(0)}, \quad
\tilde{r} = \frac{r}{b(0)}.
\label{1-110}
\end{equation}
For convenience, we hereafter drop the tilde.
As the boundary conditions, we use the constraint \eqref{1-100}
and also take into account the symmetry of the solutions to yield
\begin{equation}
	a(0) =  \frac{m^2}{4} - \frac{\varepsilon}{2}\, Q^2, \quad
	a'(0) = 0, \quad
	b(0) = 1, \quad
	b'(0) = 0.
\label{1-120}
\end{equation}

\begin{figure}[t]
\vspace{-1.2cm}
  \includegraphics[width=10.5cm]{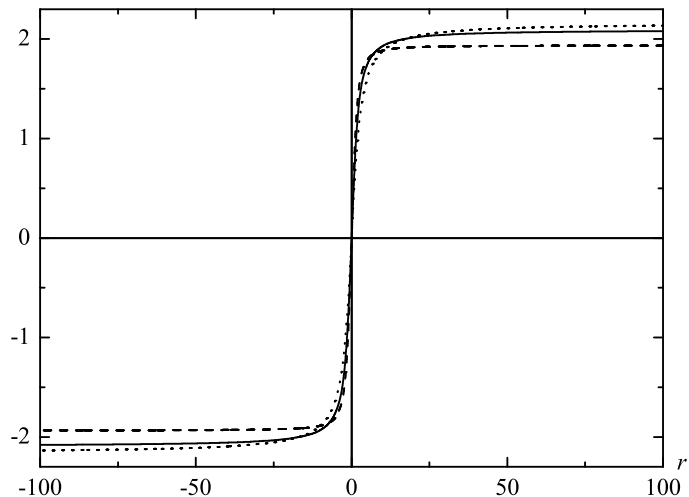}
\vspace{-1.2cm}
  \caption{The scalar function $\phi(r)$ is plotted
  for the same values of $Q$'s and $m$ as those in Fig.~\ref{arbr}.
  The curves for $Q_1, Q_2,$ and $Q_3$ are shown by the dashed, solid, and dotted lines, respectively.}
\label{phir}
\end{figure}

\section{Properties of the solutions}
\label{wh_prop}

The results of numerical calculations are shown in Figs. \ref{arbr} and \ref{phir}.
A remarkable property of the obtained solutions is that asymptotically, as $r\to \pm \infty$, the function
$b(r)$ tends to some constant value $b_\infty$.
Then, starting, say, from some $b(0) \sim {\cal O}(l_{\text{Pl}})$ and choosing the appropriate value of
$Q$ (which must be large enough, see at the end of this section), we have
$b_\infty$  of the order of $l_{\text{Pl}}$.
This assumes that the scale of the fifth dimension is always of the order of the Planck scale, i.e.,
the fifth dimension may be regarded as compactified.

According to the traditional Kaluza-Klein interpretation of a five-dimensional  metric,
we have a 4D spacetime plus a scalar field and  a
radial magnetic field: the scalar field comes from the $g_{\bar{5} \bar{5}}$ component of the 5D metric \eqref{1-50}, and
the radial magnetic field appears from the term $m \cos \theta d \varphi$. In addition, we have the massless
four-dimensional ghost scalar field $\phi$.

The line element
\begin{equation}
	dl^2 = a \left(
		d \theta ^2 + \sin^2 \theta d \varphi^2
	\right) + b^2\left(
		d \psi + m \cos \theta d\varphi
	\right)^2
\label{2-10}
\end{equation}
is the metric on the $S^3$ sphere, but this sphere is presented as the Hopf bundle
$S^3 \rightarrow S^2$
with fibre $S^1$ spanned on the coordinate $\psi$, and the metric on the base of the bundle $S^2$ is
$dl^2 = a \left(
		d \theta ^2 + \sin^2 \theta d \varphi^2
	\right) $.
Such a nontrivial bundle has the Hopf invariant defined as
\begin{equation}
	H = \frac{1}{V} \int \Theta \wedge d \Theta~,
\label{2-30}
\end{equation}
where $\Theta$ is some 1-form and $V=ab \int \sin \theta d \theta d \varphi d\psi$ is the volume of the $S^3$ sphere. In our case this form is
\begin{equation}
	\Theta = a b \left( d \psi - m \cos \theta d\varphi \right) .
\label{2-40}
\end{equation}
Substituting Eq.~\eqref{2-40} in Eq.~\eqref{2-30}, one can obtain $H = m$.
Since for the numerical calculations we chose $m=1$ then $H=1$ for the problem under consideration.

Analyzing the obtained numerical solutions, we can find their asymptotic behavior in the form
\begin{eqnarray}
	b &\approx& b_\infty + \frac{b_1}{|r|},
\label{2-60}\\
	a &\approx& r^2 + r_0^2 + a_1 \ln r^2,
\label{2-70}
\end{eqnarray}
where $b_1, a_1$ are some constants.
From the 4D point of view, the system in question contains the radial magnetic of strength
\begin{equation}
	\mathcal H_r = m\frac{b}{a}
\label{2-80}
\end{equation}
with the non-zero flux passing through the wormhole throat
\begin{equation}
	\Phi_0 =  4 \pi a(0) \mathcal H_r = 4 \pi m b(0).
\label{2-90}
\end{equation}
The presence of such non-zero flux means that asymptotic observers located on the two ends of the wormhole will see
negative and positive magnetic charges.

The numerical calculations indicate that there exists a critical value of the scalar  charge $Q_c$,
and regular solutions with the compactified fifth dimension can exist only
at $Q > Q_c$. In the case when $Q<Q_c$ there are only singular solutions.
These singular solutions are similar to those found in Ref.~\cite{Dzhunushaliev:1998ya} where no ghost scalar field is present.
As in the case considered here,
it was shown there that singularities also occur at some finite distance from the center.

\section{Conclusions}

In the present paper,
we have found static wormhole solutions within the framework of 5D Kaluza-Klein gravity
in the presence of a massless ghost four-dimensional scalar field.
The cases of the zero and non-zero Hopf invariant~$H$ (or, equivalently, of the zero and non-zero magnetic charge $m$)
have been considered.

For the case $H=0$, there exists a 5D solution with the metric function
$g_{\bar{5} \bar{5}}\equiv b^2=\text{constant}$,
which can be chosen such as to
have the size of the order of the Planck length. This means that the fifth dimension is compactified on the fibre $S^1$.
This solution is similar to the well-known 4D wormhole solution
supported by a massless ghost scalar field~\cite{wh_ghost}.

For the case of the non-zero Hopf invariant (in particular, for $H = 1$ used here), we showed the following.
\vspace{-0.2cm}
\begin{enumerate}
\itemsep=-0.2pt
\item[(1)] 
There exist regular static, spherically symmetric solutions
found numerically by solving the coupled Einstein-scalar field
equations subject to a set of appropriate boundary conditions.
The obtained solutions may be regarded as describing configurations having two types of topological nontriviality
connected with the presence of a wormhole and of a magnetic charge.
\item[(2)] The $t,r=\text{constant}$ section of the wormhole metric is topologically nontrivial,
and being the 3D sphere $S^3$ can be presented as the Hopf bundle
$S^3 \rightarrow S^2$ with fibre $S^1$.
\item[(3)] The fifth dimension is compactified. This means that the length of circle,  $2 \pi b(r)$,
spanned on the fibre $S^1$ of the Hopf bundle, is asymptotically finite: $b(\infty) \equiv b_\infty < \infty$.
In particular, $b_\infty$ can be taken of the order of the Planck length.
\item[(4)] After compactification we have the radial magnetic field whose force lines pass through the wormhole
from one asymptotic region to the other.
Thus, an asymptotic observer sees a monopole with magnetic charge equal to~$H b_\infty$.
\end{enumerate}
\vspace{-.8cm}

\section*{Acknowledgements}
We gratefully acknowledge support provided by the Volkswagen Foundation and by the grant No.~1626/GF3 in fundamental research in natural sciences
by the Ministry of Education and Science of Kazakhstan.
We also would like to thank the Carl von Ossietzky University of Oldenburg for hospitality while this work was carried out.

\end{document}